\documentclass[aps,prl,twocolumn,superscriptaddress]{revtex4}
\usepackage{graphicx}
\usepackage{amssymb}
\usepackage{natbib}

\begin{document}


\title{Rayleigh-Plateau instability causes the crown splash}


\author{Robert D. Deegan}
\email[]{rddeegan@umich.edu}
\affiliation{Department of Physics \& Center for the Study of Complex Systems, University of Michigan, Ann Arbor, MI 48109 USA}
\author{Philippe Brunet}
\affiliation{Laboratoire de M\'{e}canique de Lille, UMR CNRS 8107, Bd Paul Langevin 59655 Villeneuve d'Ascq, France}
\author{Jens Eggers}
\affiliation{Department of Mathematics, University of Bristol, Bristol BS8 1TW, UK}

\date{\today}

\begin{abstract}
The impact of a drop onto a liquid
layer and the subsequent splash has important implications for diverse physical processes such as air-sea gas transfer, cooling, and combustion.
In the {\it crown splash} parameter regime, the splash pattern is highly regular.  We focus on this case as a model for the mechanism that
leads to secondary droplets, and thus explain the drop size distribution
resulting from the splash. We show that the mean number of secondary
droplets is determined by the most unstable wavelength of the
Rayleigh-Plateau instability. Variations from this mean are governed by the width of the spectrum.
Our results for the  crown splash will provide the basis for understanding
more complicated splashes.
\end{abstract}

\pacs{47.20.Ma,47.20.Hw,47.54.De,47.80.-v}

\maketitle

The impact of a drop with a thin film of the same liquid produces a
spray of secondary droplets that results from the emission, expansion,
and breakup of one or more sheet-like jets.  Splashes are essential to diverse
physical processes and applications such as gas transfer across the
air-sea interface~\cite{ho2000}, cooling~\cite{pasandideh-fard01}, coatings~\cite{batolo06},
and combustion
~\cite{Panao05}.  The spatial pattern and
size distribution of secondary droplets are in general highly
complex ~\cite{deegan08}, varies qualitatively with
experimental conditions~\cite{worthington1897,worthington1900}, and has
yet to be understood~\cite{yarin06}.  Here we focus on the crown splash, as exemplified in Edgerton's iconic
photograph \textit{Milk Coronet}~\cite{edgerton77}, and show that number of secondary droplets is  governed by the Rayleigh-Plateau instability.  Our results can
be extended to study more complicated splashes such as when there
are multiple jets in a single impact event or the jets are irregular.

\begin{figure}[b]
\includegraphics[width=0.7\columnwidth]{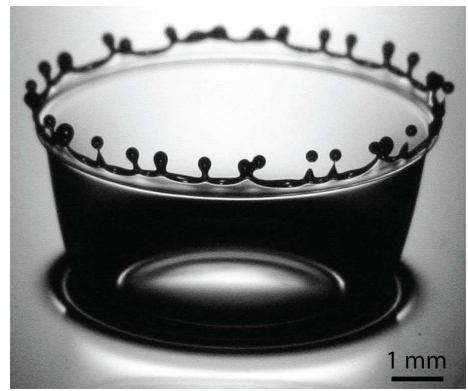}
\caption{\label{fig:crown splash} Crown splash (silicon oil: \textsf{Re}=966, \textsf{We}=874, \textsf{H$^{\ast}$}=0.2).}
\end{figure}

Figure~\ref{fig:crown splash} shows the end stage of a crown splash in
which the rim of a sheet-like jet breaks into secondary droplets
distributed almost uniformly along its perimeter. The name of the splash derives from the the resemblance of this final stage to a crown.  The events culminating in a crown splash begin with a smooth cylindrical sheet-like jet shooting outward and upward.  The leading edge of this jet is pulled by surface tension towards the sheet~\cite{Culick60,taylor59c}, and grows in diameter as it entrains fluid from the sheet.  Next, the rim develops a symmetry-breaking corrugation, and in a much later nonlinear phase of the original instability, the rim's crests sharpen into jets which  pinch off to form secondary droplets.

Due to the high speed and small scale structure of a splash there are few quantitative
time-resolved observations, and basic questions regarding the origin and evolution of splashes remain unanswered.  Leonardo da Vinci~\cite{Davinci} appears to be the first to recognize the regularity of splashes.  Worthington was the first to argue
for a surface-tension-driven instability as the origin of this
behavior~\cite{worthington1879}, yet never tested his ideas quantitatively.
The numerical computations of Rieber \& Frohn~\cite{Rieber99} support
a surface-tension-driven mechanism, whereas those of
Fullana \& Zaleski~\cite{Fullana99} argue against it.

Here we present the first experimental demonstration that the selection process is
governed by a surface-tension-driven instability.  Our experiments identified the parametric regime for crown splashes and  measured the evolution of these splashes.    A 10~cm diameter  $\lambda/10$ glass optical flat was placed on the bottom of a  12 x 13 cm glass-bottomed container. Fluid was added to the container until the optical flat was submersed, forming a film of height $h$ above the optical flat.    The orientation of the container was then adjusted so that the flat lay parallel to the fluid's surface to within $3\times 10^{-4}$ radians. The depth of the layer was varied between 150 and 300~$\mu$m depending on the fluid in order to maintain a constant ratio of drop size to layer depth. A single drop was released from a gravity fed 30 gauge needle at fixed height above the liquid layer at a rate no faster than one drop per 10~s, which ensured that the liquid layer fully relaxed between splashes. Our data on the morphology of splashes was obtained with a high speed camera (Phantom 5.0) viewing the impact from the side.  Our data for the evolution of the rim was obtained from images recorded from directly below through the glass substrates with an SLR (Canon 20D) and a single 600~ns, 6~J  pulse from a spark flash (Palflash 501, Pulse Photonics Ltd.). The flash was  triggered by the drop cutting a laser sheet focused onto a photodiode.  The triggering event was reproducible to within $\pm 5~\mu$s. The speed of the drop at impact was measured with the  high speed video camera.  The trigger was fed into a delay generator which fired the flash. By varying the delay time, the evolution of the impact was recreated from a composite
of still images.   The virtue of this technique is that it produces much higher spatial and temporal resolution than can be achieved with a high speed video camera. Our measurements of the crown splash were performed with silicon oil (density $\rho=0.92$~g/cm$^3$, surface tension $\gamma=21$~dynes/cm, dynamic viscosity $\eta$=5.2~cP).  The impact speed and drop diameter were kept constant at $U=326$~cm/s and $D=0.155$~cm, and the layer depth $h$ was either 150 or 300~$\mu$m.

\begin{figure}
\includegraphics[width=0.8\columnwidth]{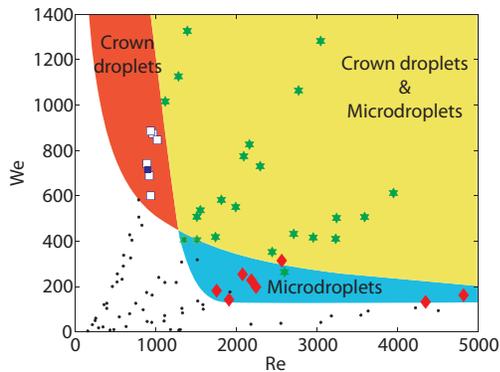}
\caption{\label{fig:phasediagram}  Qualitative character of impact for \textsf{H$^{\ast}$}=0.2.
No splash (black small circles), crown droplets with (green stars)
and without (white squares) microdroplets, and microdroplets without
crown droplets (red diamonds). The solid blue square indicates the
parameter set for all experiments reported here. A crown splash
forms exclusively in the \textsf{Crown droplets} regime.}
\end{figure}

Crown splashes occur in a limited range of experimental parameters.  The dimensionless parameters for describing droplet impact in the absence of a surrounding gas are the Weber number
$\textsf{We} = \frac{\rho D U^2}{\gamma}$,
the Reynolds number $\textsf{Re} = \frac{D U}{\nu}$, the Froude
number $\textsf{Fr} = \frac{U^2}{g D}$, and the dimensionless substrate
fluid depth $\textsf{H$^{\ast}$}=h/D$, where $g$ is the acceleration due to gravity. Past studies ignored the ambient gas, and gravity on the basis that the applicable dimensionless numbers are small.    We follow this practice here, and hence only report the $\textsf{We}$,
$\textsf{Re}$, and $\textsf{H$^{\ast}$}$.  (The recent work of Xu et al.~\cite{Xu05} found a significant influence of the ambient air on drop impact on a \emph{dry} solid.  We believe the effect of air to be much weaker in our experiment, since there is no moving contact line.) We measured the phase diagram shown in Fig.~\ref{fig:phasediagram} from a large number of experiments with different fluids, droplet diameters, and impact speeds for a fixed non-dimensional depth $\textsf{H$^{\ast}$}=0.2$. Crown splashes appear only in the regime labeled \textsf{Crown Droplets} in Fig.~\ref{fig:phasediagram}.  Outside of this domain, splashes are more irregular and complicated~\cite{deegan08}.  Our measurements of the rim evolution  correspond to dimensionless numbers $\textsf{Re}=894$, $\textsf{We}=722$, and $\textsf{H$^{\ast}$}=0.1\pm 0.02$ or $0.2 \pm 0.01$.

\begin{figure}[b]
\includegraphics[width=0.8\columnwidth]{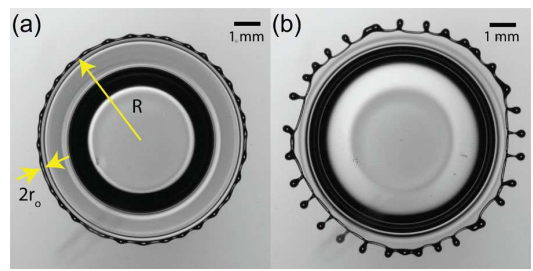}
\caption{\label{fig:numdrops} Crown splash ($\textsf{Re}=894, \textsf{We}=722$, \& \textsf{H$^{\ast}$}=0.2) from below at $t=1.85$~ms and
$t=3.15$~ms after impact showing the one-to-one correspondence between instability wavelength and the number of droplets. Yellow arrows define  rim radius $r_o$ and rim's radial distance from the impact center $R$.}
\end{figure}

\begin{figure}
  \includegraphics[width=1.0\columnwidth]{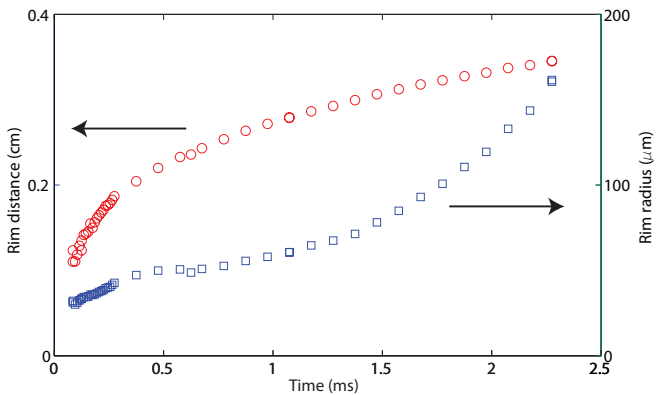}\\
  \caption{Radial distance of the rim from the impact center and radius of the rim versus time for $\textsf{Re}=894, \textsf{We}=722$, \& \textsf{H$^{\ast}$}=0.2.}\label{fig:radii}
\end{figure}

From our images, such as the example in Fig.~\ref{fig:numdrops}, we extracted the average radius of the rim $r_o$, the radial distance of the rim from the impact center $R$, and the root-mean-square amplitude and peak wavelength of the rim corrugation.  We processed the images to extract the position of the the inner and outer edge of the rim to within $\pm 6$~$\mu$m.   The average radial position for each edge was determined from a fit to a circle, and the radius of
the rim $r_o(t)$ and the radial position of the rim $R(t)$ were taken as
the half the difference and the average of these radii, respectively.  These quantities are plotted in Fig.~\ref{fig:radii}.  From the outer rim data we calculated the power spectrum, and the continuous wavelet transform $c(\theta_k,s)$, where $\theta_k$ is the translation variable and $s$ is the scale variable, using a complex Morlet wavelet.  The peak wavelength was taken to be the maximum of $c(s)=\sum_k |c(\theta_k,s)| ^2$.

\begin{figure}
\includegraphics[width=0.7\columnwidth]{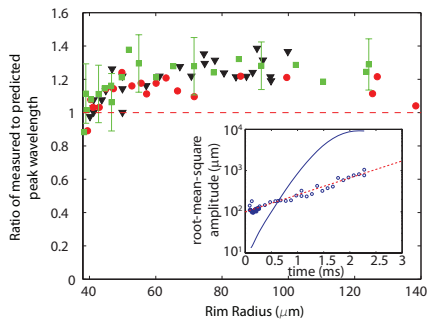}
\caption{\label{fig:peakwavelength}  Ratio of measured to theoretically expected peak wavelength versus rim radius for \textsf{Re}=894, \textsf{We}=722, \&  \textsf{H$^{\ast}$}=0.1 (triangles) or  \textsf{H$^{\ast}$}=0.2 (squares and circles).  Error bars are primarily due to the uncertainty in the measured radius of the rim.  Inset: log-linear plot of the measured root-mean-squared amplitude (blue circles), an exponential fit to this data (dashed red), and the value computer from the model from noisy initial conditions (solid blue) for $\textsf{H$^{\ast}$}=0.2$.}
\end{figure}

The peak wavelength shifts to larger values for later times which corresponds to a coarsening of the corrugation. Furthermore, the corrugation does not grow evenly at all points on the rim, but rather nucleates at several locations. These domains grow and merge, consistent with the growth of the unstable modes from random noise. At later times the corrugations sharpen and emit droplets.  The number of droplets is set by the ratio of the rim circumference to the peak wavelength, as demonstrated by the examples in Fig.~\ref{fig:numdrops}, and is hence determined by the instability.

\begin{figure}
  \includegraphics[width=0.7\columnwidth]{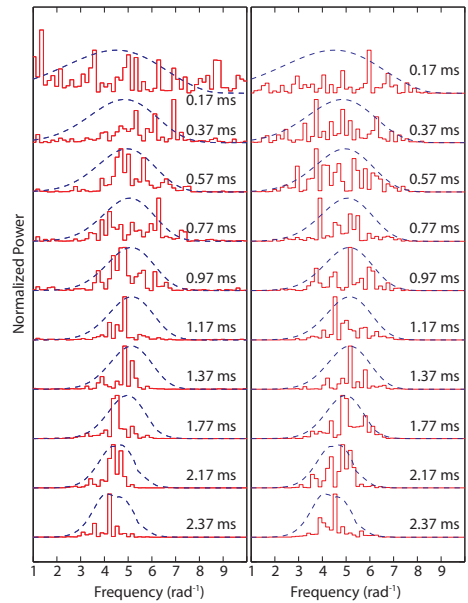}\\
  \caption{\label{fig:spectrum} Normalized power spectra versus angular frequency for various times after impact for Re=966, We=874, H=0.2.  Left: Solid (red) curve calculated from measured rim shaped.  Dashed (blue) curve calculated from Rayleigh-Plateau model using the measured rim radius and circumference.   The theoretical peak wavelength is given by the peak of the dashed curve. Right: Solid (red) curve was evolved from noisy initial conditions using the Rayleigh-Plateau model.   Dashed (blue) curve is the same as in Left.}
\end{figure}

We compared our measurements to a theoretical calculation based on the physical idea that the rim behaves like a cylinder of fluid subject to surface tension forces, ignoring the effect of the sheet attached to it. Such a fluid cylinder is susceptible to the Rayleigh-Plateau instability, which also accounts for the decay of jets into droplets~\cite{Rayleigh1878}. Its characteristic feature is a periodic variation of the rim radius with a period proportional to the average radius $r_o$.
We start at $t=0$ with a rim radius $r(\theta_j,0)=r_o+X$ defined at angular positions $\theta_j$, where $X$ is a gaussian distributed random variable.  Each of the fourier modes of the rim grows exponentially with a growth rate $\sigma_m$, which
depends on the reduced wave number $q_m$ of the perturbation, made
dimensionless with $r_0$. However, the instantaneous value of $q_m$,
and thus $\sigma_m$, changes in time according to $q_m(t) = m r_0(t)/R(t)$.
Thus the total growth of a particular amplitude is determined by the
time integral over the instantaneous growth rate obtained from
classical theory. For example, if $R(t)$ increases sufficiently,
$q$ may become smaller than one and an initially stable mode starts
to grow.  On the other hand, the stretching of the rim always {\it decreases}
the amplitude of the perturbation~\cite{Eggers08}, since longitudinal
stretching causes fluid elements to contract in the radial direction.
This leads to a reduction of the amplitude proportional to
$R(t)^{-1/2}$. The combined effect gives~\cite{eggers94}:
\begin{eqnarray*}\label{eq:computation}
    r(\theta,t)&=&r_o+\sqrt{\frac{R(0)}{R(t)}}
\sum_{m=-N}^{N} a_m \exp{\left[\imath
m \theta +\int_0^t dt'\  \sigma_m \right]}\\
 \sigma_m &=&\sigma_o\left[\sqrt{
\frac{1}{2}q_m^2(1-q_m^2)+(\beta q_m^2)^2}-\beta
q_m^2  \right]
\end{eqnarray*}
where $q_m=m\frac{r_o(t)}{R(t)}$, $\beta=\frac{3}{2}\sqrt{\ell_\nu/r_o}$,
$\sigma_o=\sqrt{\gamma/r_o^3/\rho}$, and $\ell_\nu=\nu^2 \rho/\gamma$.  The peak wavelength at a any given instant $t$, defined as the mode with the most power,  is determined by not only the instantaneously most rapidly growing mode but also by the history of the other modes.  Thus the peak wavelength is  given by the maximum of $\exp{\left\{\int_0^t dt'\  \sigma_m(t')\right\}} $.  It bears emphasizing that these calculations depend on the experimental determined variables $r_o(t)$ and $R(t)$, and that there are no adjustable parameters.

The ratio of the measured to predicted wavelength is plotted versus the rim radius in Fig.~\ref{fig:peakwavelength}.   The data shows that the wavelength is proportional to the expected value to within $\pm 20\%$ over a large variation of the rim radius, as expected for a Rayleigh-Plateau instability.  Our results also account for irregularities in the pattern. Under the action of the Rayleigh-Plateau instability, perturbations away from the most unstable wavelength are amplified as well, albeit at a smaller rate. As shown in Fig.~\ref{fig:spectrum}, the width of the spectrum of a white noise signal evolved by our equations is equivalent to that of our data.   Thus the irregularity is governed by the width of the central peak of the spectrum, a phenomenon which to our knowledge has never been considered quantitatively for the Rayleigh-Plateau instability~\cite{Eggers08}. Instead, typical measurements of linear instability impose a wavelength from the outside so as to produce a regular pattern, and irregularity is often attributed to nonlinear effects~\cite{Kadanoff00}.

The root-mean-square amplitude of the rim deformations is poorly reproduced by the our calculation, but for understandable reasons.  As shown in Fig.~\ref{fig:peakwavelength}(inset), our data is best described by an exponential growth  with a timescale of 0.59~ms for \textsf{H$^{\ast}$}=0.1 and 1.1~ms for \textsf{H$^{\ast}$}=0.2.  The expected root-mean-square amplitude, calculated from white noise initial conditions,  is not exponential--in contrast to a typical linear stability analysis--because of the time dependence of the growth rate.  On a positive note, the observed values of the growth rate fall within the range of expected instantaneous rates from the classical theory: 0.1-1.2~ms for \textsf{H$^{\ast}$}=0.2 and 0.05-0.72~ms for \textsf{H$^{\ast}$}=0.1.

We believe the deviations are a result of our simplifying assumption that the rim is cylindrical, and the unstable mode is varicose. First, even if the sheet is thin compared to the rim radius, the attachment constrains the evolution of the rim and the reduction of surface energy due to corrugation is not as great as it would be for an unconstrained cylinder. A model calculation in which the curve along which the sheet attaches to the rim is kept straight, as observed experimentally, shows that the growth rate decreases.  Second, the gradual transition from rim to sheet will slow the growth of the instability. For example, in the extreme case in which the rim is simply a semicircular end of the sheet, there no surface energy to be gained from a local reduction of the rim radius, and thus no instability at all.

We also considered and rejected other proposed mechanisms to explain the symmetry breaking
of the crown.  Bremond and Villermaux~\cite{bremond06} suggested a Rayleigh-Taylor instability stemming from the deceleration of the rim.  Our measurements of the rim's deceleration show that the predicted wavelength is 2 times greater at the earliest times and 8 greater at late times.  Gueyffier \& Zaleski proposed a
Richtmyer-Meshkov type instability due to the sudden deceleration
during impact~\cite{Gueyffier98}.   Our measurements show that the instability grows exponentially  which is inconsistent with the slower linear growth of a Richtmyer-Meshkov instability. Yarin proposed a nonlinear amplification mechanism~\cite{yarin95,yarin06}. This mechanism does not select a particular wavelength in contradiction to our measurements which clearly show wavelength selection.

In conclusion, we have demonstrated for the first time the selection
of a preferred wavelength that sets the number of secondary droplets
resulting from a crown splash. Its number is proportional to the
circumference of the splash sheet, divided by the radius of the rim.
The irregularity of the crown is set by amplification of random
noise by the spectrum of growth rates of the Rayleigh-Plateau instability.
The similarity of the splash patterns for other parameter regimes suggests
the same mechanism to be at work, but with a broader range of initial
perturbations.

\begin{acknowledgments}  We thank Daniel Bonn for comments on the manuscript, and Christophe Josserand for discussions.
\end{acknowledgments}


\end{document}